# DeePAW: A universal machine learning model for orbital-free ab initio calculations

Tianhao Su[1,‡], Shunbo Hu[1,2,‡], Yue Wu[1,‡], Runhai Oyang[1], Xitao Wang[1], Musen Li[1,*], Jeffrey Reimers[3,*], Tong-Yi Zhang[1,4,*]

[1] Materials Genome Institute, Shanghai Engineering Research Center for Integrated Circuits and Advanced Display Materials, Shanghai Frontier Science Center of Mechan informatics, State Key Laboratory of Materials for Advanced Nuclear Energy, Shanghai University, Shanghai 200444, China

[2] Institute for the Conservation of Cultural Heritage, Key Laboratory of Silicate Cultural Relics Conservation (Ministry of Education), Shanghai University, Shanghai 200444, China

[3] School of Mathematical and Physical Sciences, University of Technology Sydney, Ultimo, New South Wales 2007, Australia

[4] Guangzhou Municipal Key Laboratory of Materials Informatics, Sustainable Energy and Environment Thrust, Advanced Materials Thrust, Hong Kong University of Science and Technology (Guangzhou), Guangzhou 511400, Guangdong, China

‡ These authors contributed equally to this work.

* Corresponding authors.

**Abstract**

Developing universal machine learning models for ab initio calculations is the frontier of materials cutting edge research in the new era of artificial intelligence. Here, we present the Deep Augment Way model (DeePAW) that is a universal machine learning (ML) model for orbital-free (OF) ab initio calculations, based on the density functional theory (DFT). DeePAW is currently the best OFDFT ML model according to the three criterions, 1) covering the largest number of elements, 2) having the widest application capability to diverse crystal structures, and 3) achieving the highest prediction accuracy without further fine-tuning. These scientific merits and innovations of DeePAW are stemmed from the novel SE(3)-equivariant double massage passing neuron networks. Besides predicting electron density distributions, DeePAW predicts formation energies of crystals as well and therefore paves an efficient avenue for multiscale materials modeling beyond conventional electronic structure calculation methods.

**Introduction**

The fast development of artificial intelligence (AI) has been greatly advancing the field of ab-initio calculations. Roughly, the AI ab-initio calculations can be divided into three areas, namely machine learning interatomic potentials (MLIP)[1], deep density functional theory (DFT) Hamiltonian models[2], and AI electron density models for orbital-free DFT (OFDFT) calculations to replace the expensive orbital Kohn–Sham (KS) DFT calculations. Considerable efforts have been devoted to the development of MLIP[3,4]. In particular, many MLIPs exhibit great universal when they were trained on large materials datasets, such as Materials Project[5], OMat24[6], and Alexandria[7], with each dataset covering over 80 elements. Examples are the recent CHGNet[8], MatterSim[9] and UMA[10] and the top three models, eSEN-30M-OAM[11], EquFlash[12], and Nequip-OAM-XL[13], on the leaderboard Matbench Discovery[4,14] that evaluates machine learning crystal stability predictions. There are also some MLIPs, such as Deep Potential (DP)[15] and NeuroEvolution Potential (NEP)[16], which are seamlessly integrated with the conventional MD simulation code, LAMMPS, to conduct large scale simulations. Deep Potential (DP)[15] is usually trained on a special dataset, which is built up on many KSDFT calculations done by users themselves. NEP[16] covers 16 metallic elements and thus is relatively universal. In general, DP accuracy is higher than NEP accuracy, while the MD simulation speed of NEP is faster than that of DP. Nevertheless, these MLIPs have tremendously facilitated MD simulations at large-length and long-time scales to investigate diffusion, phase transformation, microstructure evolution, and associated various material behaviors and properties[3]. In deep Hamiltonian learning, most models fucus on particular chemical systems, e.g., eSCN [17], QHNet[18], and TraceGrad[19], DeepH[20], DeepH-E3[21], etc., and all of them have shown to be powerful AI-Hamiltonians. DeepH and its late variants [20-23] are based on Message Passing Neural Network (MPNN)[24] and these variants[25-27] incorporate the e3nn library[28] into the DeepH framework. Particularly, DeepH-2 model is universal, trained on 12,062 structures, which are nonmagnetic and contain the atom number in a unit cell less than 150 atoms in the Materials Project (MP) database[5], and covers 24 elements. HamGNN is a universal deep Hamiltonian model, trained on a larger dataset of ~55,000 structures covering 75 elements and demonstrates superior performance with a mean absolute error (MAE) of 5.4 meV/atom[29]. NextHAM further advances the field by covering 67 elements and achieving a lower MAE of 1.4 meV/atom[30].

Deep learning electron density models provide an alternate promising route for electronic structure predictions based on OFDFT and ground-state materials properties are expressed by the minimal internal energy of electron kinetic energy and potential energy of studied structures. As described in the overreviews[31,32], most the AI-OFDFT models[33-35], reported in the literature so far, are on molecules and special chemical compounds with few elements. For examples, DeepDFT is tested one three datasets involving only 9 elements in total[33], which are 1) QM9 dataset containing 134k small molecules with each up to nine heavy atoms of C, N, O, or F, and some H atoms, 2) liquid carbonate electrolyte ($C_3H_4O_3$) dataset and 3) $Li_xNi_yMn_zCo_{(1-y-z)}O_2$ lithium ion battery cathodes (NMC) dataset. JLCDM[34] predicts the electron densities of molecules and solids, including benzene ($C_6H_6$), Al, Mo, and two-dimensional (2D) $MoS_2$[34]. Few universal AI-OFDFT models are also developed. For instance, ChargE3Net[35] is trained on a massive dataset of over $10^5$ structures over 80 elements in the MP database and captures the complexity and variability in the data, leading to a significant 26.7% reduction in self-consistent iterations when used to initialize DFT calculations on unseen materials. The performance accuracy of ChargE3Net[35] is gauged by a normalized mean absolute percentage error (NMAPE) of 0.523% in comparison with those obtained directly from KSDFT and the maximum NMAPE for individual elements can reach 3.0%[35]. Recently, EAC-Net leverages the MP database to develop EAC-mp[36], a universal charge density predictor that couples atoms and grids to integrate both strengths of grid-based and basis-function based frameworks. EAC-mp[36] covers 84 elements and demonstrates exceptional performance with NMAPE below 1% across diverse material systems. Although the aforementioned works demonstrate the growing opportunities of AI-OFDFT for materials predictions in the era of AI, no accurate universal AI-OFDFT models have been reported yet, which stimulate us to develop a universal AI-OFDFT model.

Here, we introduce the novel Deep Augmented Way (DeePAW) method for OFDFT electron density functional and formation energy prediction of crystalline structures. Inspired by the Projector Augmented-Wave (PAW) method[37], DeePAW partitions the electron density functional into a smooth part and a fluctuated part. Equipped with the e3nn [28], DeePAW adopts a double-MPNN architecture to ensure that the consistent learning between atomic configuration and electron density becomes a robust, universal and strongly powerful model at electron density functional and formation energy predictions after trained on the MP database. Without any fine tuning, the pretrained model shows highly accurate predictions of formation energy and electron density functional across 3D perfect and disordered 3D structures inside and outside the MP database, ferroelectricity, 2D and 1D materials, catalyst, and further photon absorption spectra when extending to time dependent (TD) DeePAW.

## Results

**DeePAW architecture**

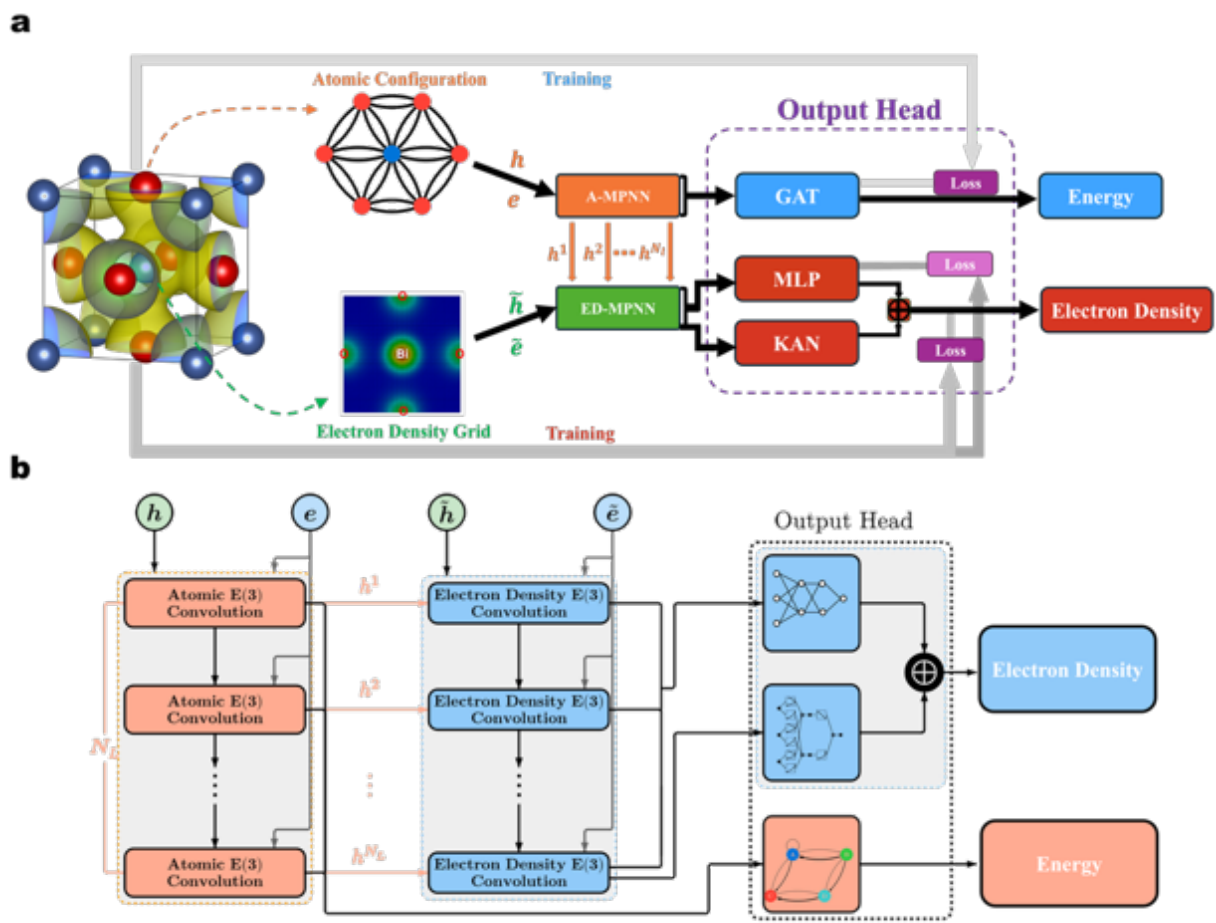

**Figure 1 a.** Schematic illustration of the workflow and model architecture of DeePAW. The input crystal can be one in the training dataset, with labeled energy and electron density functional that are fed into corresponding loss functions, or an unseen one, which energy and electron density will be predicted. The crystal information, including all involved atomic number and atom sites and lattice vectors, is used to construct atomic configuration and electron density grid. The atomic configuration is represented by atomic embeddings $h$ and edge embeddings $e$, while the electron density grid is represented by grid point embeddings $\tilde{h}$ and edge embeddings $\tilde{e}$. The atomic embeddings are updated purely within its own atomic MPNN, whereas the grid point embeddings are updated through its electron density MPNN, layer-by-layer, and with the input of atomic embeddings at the same level of layers. The output head includes three networks: Graph Attention Networks (GAT)[38] for formation energy output, and MultiLayer Perceptrons (MLPs) and Kolmogorov Arnold Networks (KANs)[39] for electron density output. **b**, Both atomic MPNN and electron density MPNNs are built up by using the e3nn library[40] to implement E(3)-equivariant convolution. The atomic MPNN passes the atomic embeddings, layer by layer, to the corresponding electron density MPNN and there are $N_L$ layers in each of the double MPNNs.

**Fig. 1a** shows the e3nn-based DeePAW architecture, which is built-up by double graph massage passing neural networks (MPNN) and one output head including three modules. Using the e3nn library[40], a calculated crystal is represented by one conventional crystal graph, called atomic graph, with the atomic nuclei of the crystal as nodes $v_i (i = 1, 2, \cdots, N_a)$ and the connections between atomic nuclei as edges $e_{ij}$, where $N_a$ denotes that number of total atoms in the crystal. The atomic configuration provides the electron space around atomic nuclei, where the electron space is represented by electron density graph, where the nodes are the 3D grid points, $r_k (k = 1, 2, \cdots, N_\rho)$, where $N_\rho$ denotes that number of total grid points in a calculated lattice cell, and edges are defined as the connections of a grid point to its connected nuclei. The electron density $\rho(r)$ functional is discretized in the numerical calculations with $\rho(r) \geq 0$ and $N_e = \int \rho(r) dr = \sum_i^{N_\rho} \rho(r_i)$, where $N_e$ is the number of total valence electrons in the calculated many-electron crystal. There are $N_L$ layers in each MPNN, where $N_L$ is a hyperparameter and its optimal value $N_L = 3$ is determined by the ablation test. The node embeddings are updated layer-by-layer via the message passing (MP) E(3)-equivariant convolution and at each layer only the updated node embeddings of the atomic MPNN are fed into the electron density MPNN. The e3nn library[40] is used in the double MPNNs with period boundary conditions. With the atomic configuration of a crystal lattice, the initial node features are embedded on the atomic number, and the edge features are initially embedded by the Gaussian basis function (GBF) (see Methods for detail) and fed into every layer of atomic MPNN, as shown in Fig. 1b. The spherical harmonics embeddings in terms of angular momentum quantum number and magnetic quantum number appear in the all layers of MPNN (see Methods for detail). For the electron density MPNN, the initial node features might be homogeneously or randomly set-up under the constraints $\rho(r) \geq 0$ and $N_e = \sum_i^{N_\rho} \rho(r_i)$. The initial edge features also take the GBF embeddings, where the distance is between a grid point to its connected nucleus, and the spherical harmonics embeddings at that grid point and the node embeddings appear in all layers (see Methods for detail). The atomic MPNN and the electron density MPNN deliver all the node embeddings of their $N_L$ layers to the GAT (see the Supplementary Materials for brief description) and MLPs heads, respectively, while the last layer in electron density MPNN feeds its output into the KAN (see the Supplementary Materials for brief description) head. The GAT head outputs the formation energy and both MLPs and KAN heads jointly predict electron density functional (see Methods for detail). The MLPs head delivers the smoothly densities $\tilde{\rho}^{MLP}(r)$ at all grid points, similar to the pseudo wavefunctions in the classical PAW[37] (see the Supplementary Materials for brief description). The true electron densities $\rho(r)$ at all grid points are taken from the MP database and used to calculate the residues of electron densities $\Delta\rho(r)$ between $\rho(r)$ and $\tilde{\rho}^{MLP}(r)$. The KANs take the node embeddings of the last layer in the electron density MPNN as input features and deliver the predicted residues of electron densities $\Delta\tilde{\rho}(r)$, where the sum of $\tilde{\rho}^{MLP}(r) + \Delta\tilde{\rho}(r) = \tilde{\rho}(r)$, as the predicted electron density, is fed, with the labels $\rho(r)$, into the loss function of mean absolute error to train the model (see Methods for detail). In this sense, the KANs behave like all-electrons wavefunctions in the classical PAW[37]. That is why this universal machine learning model is named the DeePAW.

**General Performance of DeePAW**

DeepPAW is trained on the MP database, which includes 117,452 crystal structures, spans of all seven crystal systems and 212 space groups, and covers 88 elements after excluding the two elements of He and Ne, because of He or/and Ne being involved only three crystals (see Datasets for detail). The general performance of DeePAW on the testing set in the MP database is evaluated by Normalized Mean Absolute Percentage Error (NMAPE), defined by $\varepsilon_{nmape} = \frac{\sum_i^{N_\rho^{test}} |\rho(r_i) - \tilde{\rho}(r_i)|}{\sum_i^{N_\rho^{test}} |\rho(r_i)|}$, where $N_\rho^{test}$ denotes the number of grid points in one minibatch, $\rho(r_i)$ and $\tilde{\rho}(r_i)$ are the true and predicted charge densities, respectively. Statistically analyzing the prediction results of many minibatches yields the mean $\varepsilon_{nmape}$ value and standard deviation. **Table 1** lists the general performance of DeePAW, ChargE3Net[35], DeepDFT[33] and EAC-mp[36], all trained and tested on the MP database, where the $\varepsilon_{nmape}$ values of the other three models are copied from their original publications.

Table 1. The general performance on electron density of ChargE3Net[35], DeepDFT[35], EAC-mp[36] and DeePAW.

| Model | ChargE3Net[35] | DeepDFT[33] | EAC-mp[36] | DeePAW |
|---|---|---|---|---|
| NMAPE % | 0.523 ± 0.010 | 0.799 ± 0.010 | 0.71 ± 0.23 | 0.351 ± 0.002 |
| # of elements | 80 | 9 | 84 | 88 |
| formation energy | no | no | no | yes |

Obviously, DeePAW performs superior than ChargE3Net, DeepDFT and EAC-mp in the prediction of electron density, in addition covering more elements than them and providing formation energy.

Table 2 lists the general performance on electron density and energy of DeePAW on the seven crystal systems in the test set, indicating that DeePAW has achieved the state-of-the-art (SOTA) accuracy in predicting electron density distributions and formation energy across the seven crystal systems.

Table 2. The general prediction performance of DeePAW on electron density and formation energy of DeePAW in the seven crystal systems of the testing set.

| Crystal System | Electron Density (NMAPE %) | Energy (MAE eV/atom) |
|---|---|---|
| Cubic | 0.26 ± 0.001 | 0.047 ± 0.002 |
| Trigonal | 0.33 ± 0.001 | 0.045 ± 0.002 |
| Tetragonal | 0.38 ± 0.001 | 0.047 ± 0.002 |
| Orthorhombic | 0.41 ± 0.001 | 0.044 ± 0.002 |
| Hexagonal | 0.41 ± 0.001 | 0.045 ± 0.002 |
| Monoclinic | 0.43 ± 0.001 | 0.042 ± 0.002 |
| Triclinic | 0.45 ± 0.001 | 0.040 ± 0.002 |

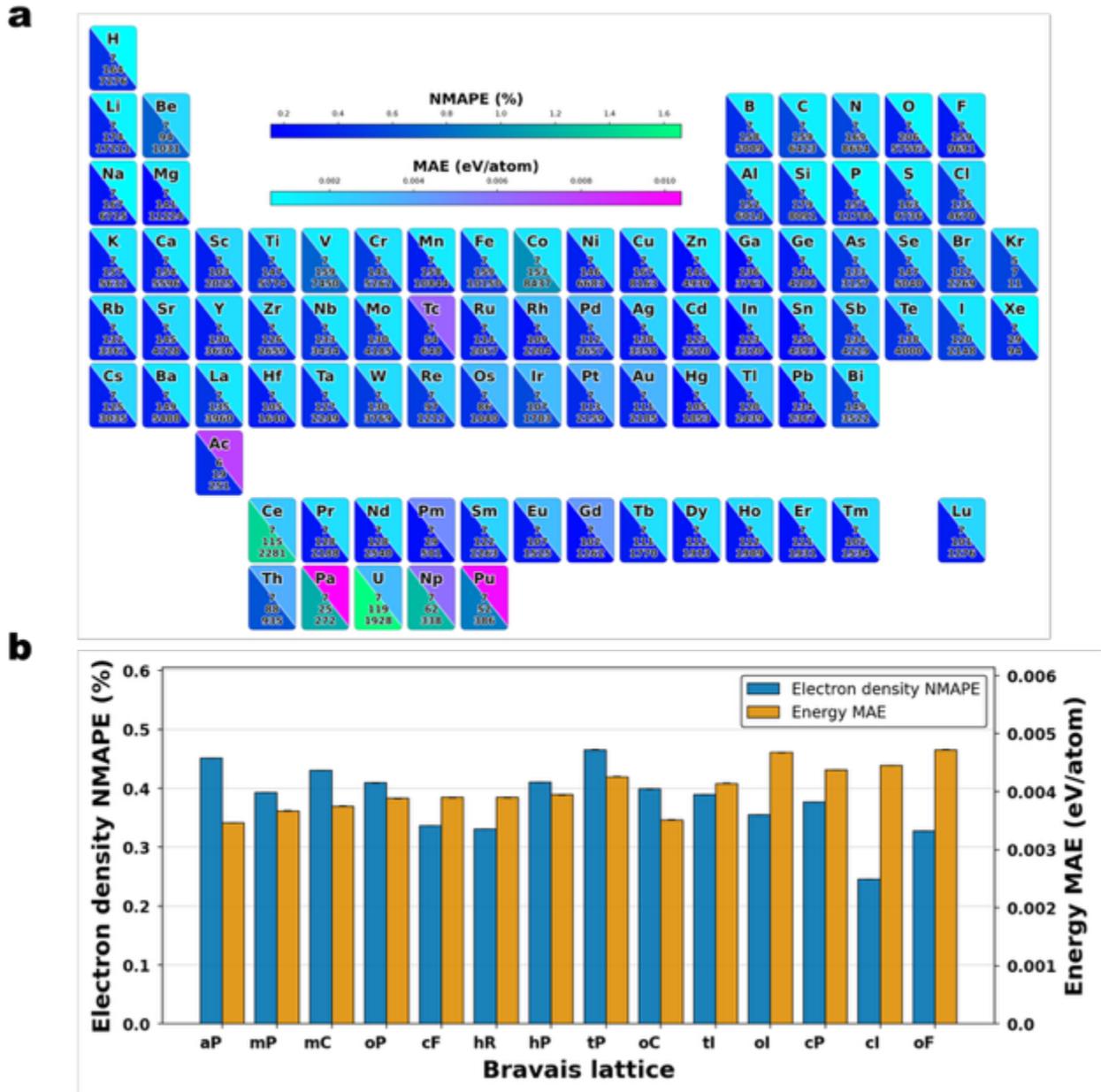

**Figure 2 a.** The DeePAW performance on each of 88 elements in the MP crystal structures, where the colors of upper and lower triangles for each element denote the mean absolute error (MAE) for formation energy and the $\varepsilon_{nmape}$ value for electron density, respectively, and the three numbers denote each element participating numbers in seven crystal systems, 212 space groups and 117,452 crystal structures, respectively. **b**. The DeePAW performance on the 14 Bravais crystal lattices in the MP crystal structures, where yellow and blue colors represent the $\varepsilon_{nmape}$ values for electron density and energy, respectively, and aP, mP, mS, oP, oS, oI, oF, tP, tI, hR, hP, cP, cI, and cF denote the triclinic, monoclinic primitive, monoclinic base-centered, orthorhombic primitive, orthorhombic base-centered, orthorhombic body-centered, orthorhombic face-centered, tetragonal primitive, tetragonal base-centered, rhombohedral primitive, hexagonal primitive, cubic primitive, cubic bay-centered, and cubic face-centered, respectively.

Figure **2a** is a periodic table, showing the following information of the 88 elements. In each element square, the colors of upper and lower triangles denote the MAE (eV/atom) value for formation energy and the $\varepsilon_{nmape}$ value for electron density, respectively, and the three numbers below the element symbol denote the element-participating numbers in seven crystal systems, 212 space groups and 117,452 crystal structures, respectively. For example, for element V, the azure color of lower triangle and the sky-blue color of upper triangle indicate that its MAE and $\varepsilon_{nmape}$ values for formation energy and electron density are about 0.00154 eV/atom and 0.735 %, respectively, and the three numbers of 7, 159 and 7,450 indicate that V is a chemical component in all seven crystal systems, but only in 159 space groups and 7,450 crystal structures. The general performance of DeePAW is extremely outstanding, across the periodic table and crystallographic systems. The element-wise analysis of Fig. 2**a** reveals heterogeneous prediction errors, with normalized mean absolute percentage error (NMAPE) for electron density and mean absolute error (MAE) for formation energy varying systematically across chemical species. Some elements, like uranium, neptunium, and plutonium, have the highest electron density prediction errors (NMAPE ~0.8–1.6%), which are accompanied by the elevated formation energy MAEs (~0.003–0.007 eV/atom). This relatively poor performance of DeePAW might reflect the great scattering and large uncertainty in the KSDFT dataset, because it is difficult to accurately calculate the complex electronic structure of f-electron systems, which include strong electron correlation effects and spin-orbit coupling. Transition metals, such as cobalt, vanadium, and cerium, exhibit intermediate errors, whereas main-group elements generally yield substantially lower deviations. Inert gases display minimal prediction errors, with argon achieving the lowest metrics among elements with statistically significant representation in the dataset. Fig. 2b displays the systematic analysis across the 14

Bravais lattice types reveals electron density NMAPE spanning 0.24% (body-centered cubic, cI) to 0.46% (base-centered monoclinic, mC, and primitive orthorhombic, oP), with error bars of ±0.001, which is too small to show in Fig. 2**b**. Accordingly, the formation energy MAE ranges from 0.0034 to 0.0048 eV/atom (±0.00003 eV/atom uncertainty). High-symmetry cubic structures (cI, face-centered cubic cF, and primitive cubic cP) consistently achieve superior accuracy relative to lower-symmetry monoclinic and orthorhombic systems. This performance hierarchy correlates with the degree of crystallographic symmetry, suggesting that the model more effectively captures the electronic structure of highly symmetric environments.

**Performance of DeePAW on three-dimensional (3D) crystals**

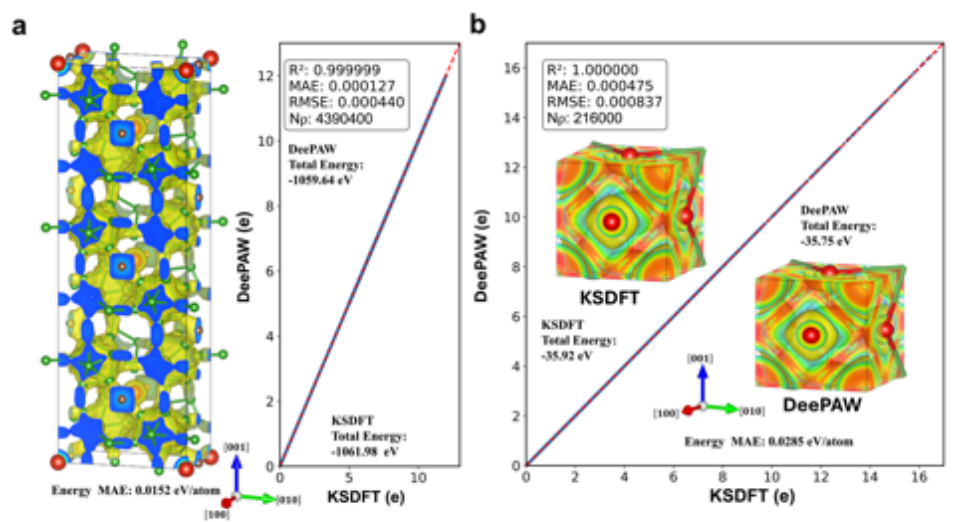

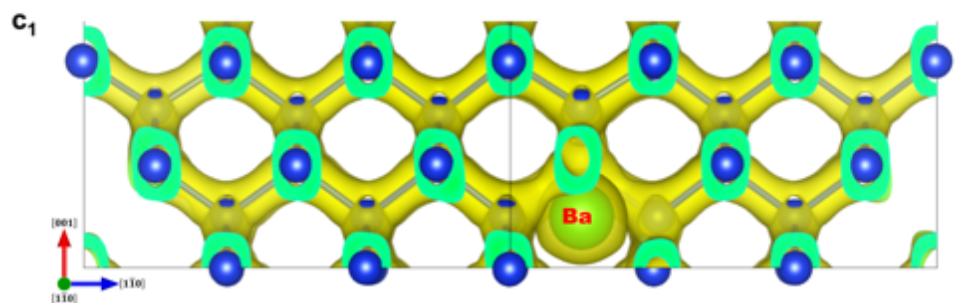

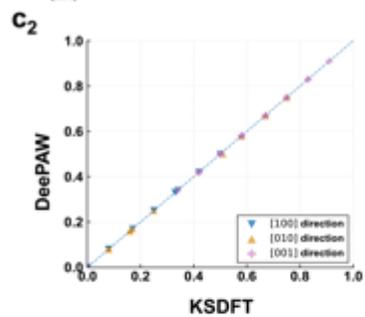
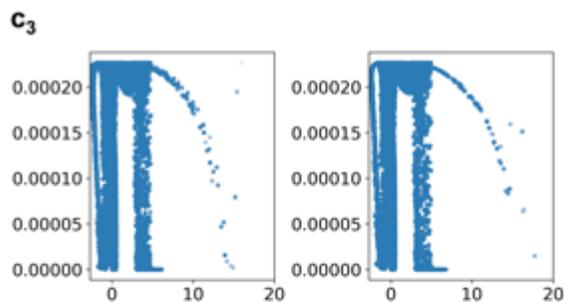

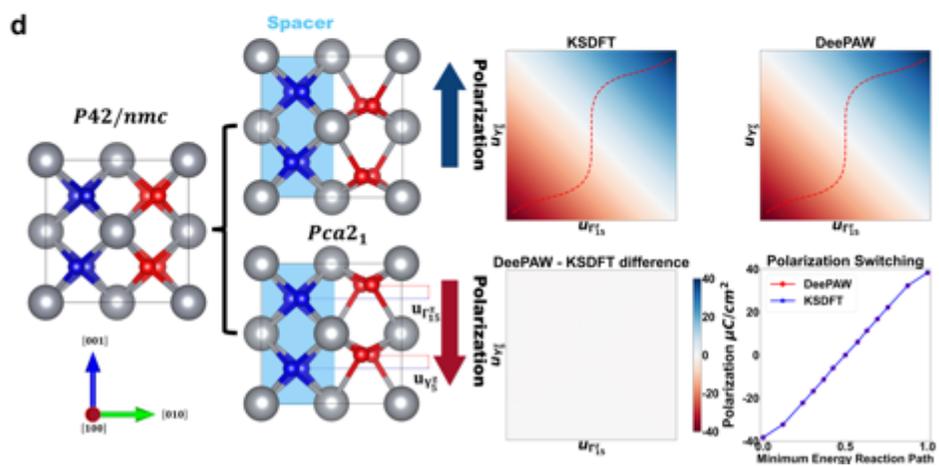

**Figure 3 a**. The $V_4C_6B_{144}$ crystal is the largest number of atoms per unit cell (154 atoms per cell) in the test dataset and the predicted formation energy and electron density are in excellent agreement with the corresponding labels. **b**. The scatter plot of predicted electron density versus the KSDFT calculated ones for $Pm\bar{3}m$ LaGaO₃ crystal[41], in which the upper-left and lower-right insets show, respectively, the calculated and predicted DORI analysis and formation energy value, indicating the perfect agreement. $c_1$, Real-space visualization of electron density distribution in the 2x2x1 Si superlattice with an interstitial Ba atom (217 atoms in total), selected from the MPcontribs dataset[42]. $c_2$ The site-wise predictions of lower energies as a function of the distance from the defect site (in fractional coordinates) along the three crystallographic axes. $c_3$, The plot of DORI versus $\lambda_2$ for DeePAW predicted is in perfect agreement with that for KSDFT calculated, indicating the great power of DeePAW on the electron density prediction at defects. **d**, Ferroelectric switching in HfO₂, comparing the DeePAW predictions and the KSDFT calculated, where polar phonons $\Gamma_{15}^z$ and antipolar phonons $Y_5^z$ vary in the same range from -u to u. The white (zero) difference between DeePAW predictions and the KSDFT calculated in the entire polar and antipolar phonon space demonstrates the great power in the ferroelectric switching prediction.

As examples, **Fig. 3** presents the performance of DeePAW in several representative 3D crystals. **Fig. 3a** shows the DeePAW prediction on the electron density of the crystal $V_4C_6B_{144}$ (space group $P222$) containing the largest number of 154 atoms per unit cell in the testing set of the MP database, which corresponds a large number of grid points ($N_\rho$), 4,390,400, in the testing set. As expected, the predicted formation energy is identical to the labeled and the electron density prediction on the structure has extremely high prediction accuracy with the coefficient of determination $R^2 = 0.999999$, the mean absolute error, MAE= 0.000127 and the root-mean-square error RMSE= 0.000440. DeePAW exhibits the same excellent performance on the charge density of crystals in other databases. For instance, a randomly selected test cubic structure LaGaO₃ (space group $Pm\bar{3}m$) from the perovskite-5 dataset[41] is taken to illustrate the prediction power of DeePAW. The lattice cell of LaGaO₃ contains 5 atoms and corresponding $N_\rho$=216,000 grid points. The high prediction accuracy on formation energy is evidenced by the almost identical vales and on electron density is evidenced by $R^2 \approx 1$, MAE= 0.000475 eV/atom and RMSE= 0.000837 eV/atom. In addition to exceptional prediction accuracy, the DORI (Density Overlap Region Index) analysis is conduced and shown in the upper-left and lower-right insets of **Fig. 3b** for KSDFT and DeePAW, respectively. The DORI texture from DeePAW and that from KSDFT exhibit remarkable agreement with each other in their geometric patterns, indicating that DeePAW not only achieves high prediction accuracy in electron density, but also faithfully reproduces the topological features of electron density at a high level of geometric fidelity.

Besides the perfect crystals, we apply DeePAW to discover the potentially favorable interstitial sites induced by substitute atoms, because atoms, like H, Li, C, O, etc., with small size prefer to occupy the interstitial sites with lower energy. An electron-density-based general cation insertion algorithm[43] is proposed to discover Li-ion cathode materials, which is based on the one-to-one correlation between the formation energy of interstitial atom and the local minimum of charge density in the host crystal. Physically, these local minima within the electron density field correspond to spatial sites with reduced electrostatic repulsion and therefore the formation energy for Li insertion is correspondingly low. Since DeePAW predicts the electron density directly, it is much more convenient to apply DeePAW to discover these local minima within the electron density field. To demonstrate the DeePAW prediction power in this aspect, we randomly select a 2x2x1 superlattice diamond structure silicon (space group $P\bar{4}3m$) including one Ba atom at the tetrahedron interstitial site (217 atoms in total) from the MPcontribs dataset[42] (see Methods for detail) and feed the structure directly into DeePAW. The algorithm to calculate minimal electron densities is the charge-density-driven computational protocol implemented within the Pymatgen package[44]. The KSDFT calculation is conducted later to verify the DeePAW discovery. **Fig. 3c$_1$** shows the visualization of electron density in the point-defect Ba doped Si structure and Fig. 3**c$_2$** illustrates the local minimal electron density of DeePAW predicted versus that KSDFT calculated, showing highly accurate predictions with $R^2$ about 1. Figure 3**c$_3$** presents the DORI-$\lambda_2$ two-dimensional maps generated from both KSDFT and DeePAW. The electron density prediction accuracy, remaining extremely high with $R^2$=0.999 and NMAPE $\varepsilon_{nmape}$ =0.20. Furthermore, error analysis reveals that the prediction errors in the defective system are all extremely low at the level $10^{-14}$ of mean absolute error at every $\lambda_2$ value, thereby indicating without error clustering in the high-gradient bonding regions ($\lambda_2 < 0$). By comparing DeePAW-discovered local minima with those KSDFT-computed, we can claim that DeePAW reconstructs not only the global electron density functional but also its fine-grained features, such as gradient variations and potential basins around defects. This is especially important in doped or defective systems, where small perturbations in electron density can lead to significant changes in ionic transport pathways and defect interactions.

Lee et al.[45] discovers the distinct ferroelectricity in orthorhombic (Pca2$_1$) hafnium dioxide (HfO$_2$), where the electric dipoles are extremely localized within its irreducible half-unit cell widths (~3 angstroms) and individually switchable without creating any domain-wall energy cost. DeePAW predicts the electron density with high accuracy and thereby is ideal in handling the ferroelectricity in HfO$_2$, where the center of positive charge deviates from the center of negative charge. Especially, no ferroelectric crystals can be found in the used MP database and therefore the investigation of ferroelectricity in orthorhombic HfO$_2$ will test the generalization power of DeePAW. **Fig. 3d** shows the half spacer cell and the half ferroelectric cell of an orthorhombic HfO$_2$ lattice cell and the polarization strength of ferroelectricity depends on polar phonons $\Gamma_{15}^z$ and antipolar phonons $Y_5^z$, where silver spheres denote Hf atoms; red and blue spheres indicate oxygen atoms in the ferroelectric layer with up and down polarization, respectively; and green spheres indicate oxygen atoms belonging to the spacer layer. The displacements associated with the phonons are denoted by $u_{\Gamma_{15}^z}$ and $u_{Y_5^z}$ in the range from -u to u, which magnitude depends on temperature. When the polar and antipolar phonons condense in-phase with equal magnitude to generate an orthorhombic structure that consists of alternating spacer layers and ferroelectric layers with up (top), and down (bottom) polarization, respectively. The $u_{\Gamma_{15}^z}$ and $u_{Y_5^z}$ space is grided into 20x20 points and at every point the polarization strength is calculated by DeePAW and by KSDFT for verification. The results indicate that the polarization strength calculated by DeePAW agree completely with that from KSDFT, as shown in Fig. 3d, where the dashed curve denotes the path of polarization switch[45] and the last small figure gives the switch detail. It is no doubt that DeePAW trained on the MP database has the great power to predict the electron density quickly in ~ 1 seconds/item, while the self-consistent KSDFT takes ~ 100 seconds/item.

# Performance of DeePAW on low-dimensional crystals

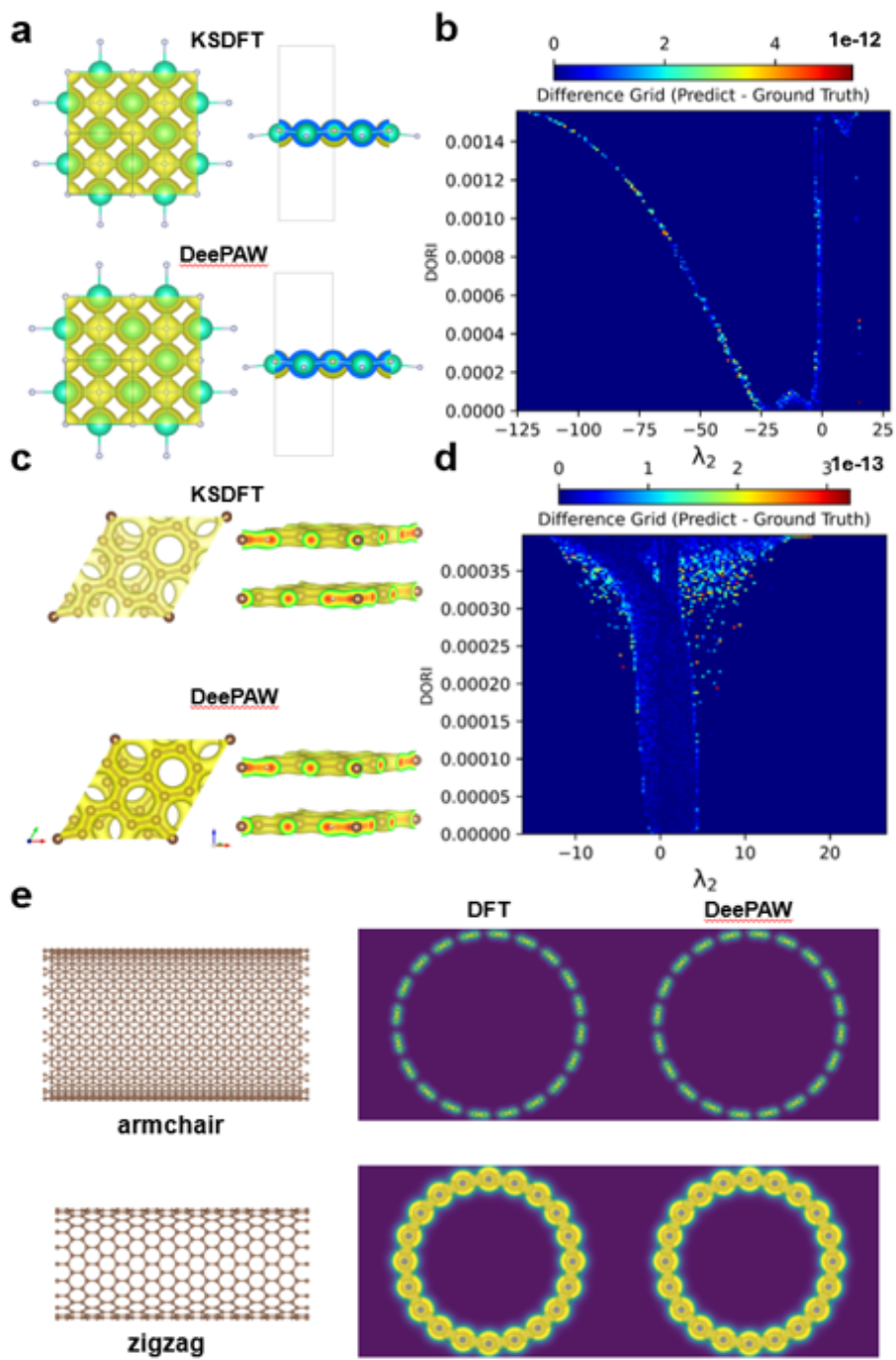

**Figure 4.** Prediction results of DeePaw on 2D and 1D materials. **a.** The predictions of electron density and formation energy of the monolayer 2D CsF structure, where blue and white denote the Cs and F atoms, respectively, and **b.** corresponding DORI-$\lambda_2$ analysis. **c**. The predictions of electron density and formation energy of twist bilayer graphene (2, 3) in twisted bilayer graphene and **d.** associated DORI-$\lambda_2$ analysis. **e.** The visualization of predicted electron density distributions for armchair and zigzag carbon nanotubes and associated formation energies.

DeePAW not only exhibits considerable capacity in the prediction of electron density and formation energy in 3D crystals, but has the great power in the prediction of electron density and formation energy in 2D and 1D crystals. Fig. 4a and Fig. 4b show the electron density distribution and formation energy, and associated DORI-$\lambda_2$ analysis on the randomly sampled monolayer 2D material CsF structure (space group P4/nmm) from the C2DB dataset (see Methods for detail), where the 2D lattice contains 4 atoms with 1,975,680 grid points. DeePAW achieves high prediction confidence in both electron density and formation energy with NMAPE=0.22 in the electron density prediction and with $R^2$ = 0.9993 in the DORI-$\lambda_2$ analysis. DeepH[20] studies twisted 2D materials with an arbitrary twist angle $\theta$ and hence builds a twisted materials dataset. The twisted angle $\theta$ is denoted by $(m, n)$ via $\cos(\theta) = \frac{n^2+4nm+m^2}{2(n^2+nm+m^2)}$, when the twist changes a lattice vector **V**$(m, n)$ in one layer to **V**′$(n, m)$ in another layer with the $(m, n)$ coordinates with respect to the basis vectors $\boldsymbol{a}\left(\frac{\sqrt{3}}{2}, -\frac{1}{2}\right)$ and $\boldsymbol{b}\left(\frac{\sqrt{3}}{2}, \frac{1}{2}\right)$[46]. We randomly select one twisted bilayer graphene (2,3) from the dataset[20] (see Methods for detail), where the twisted bilayer graphene contains 32 carbon atoms, as shown in Fig. 4c. Notably, the DeePAW results, validated by KSDFT, demonstrate that DeePAW can accurately capture the complex electron redistribution within such twisted bilayer graphene. Fig. 4d shows the difference in DORI-$\lambda_2$ analysis calculated by DeePAW and KSDFT. The interlayer van der Waals regions in the bilayer graphene are characterized by $\lambda_2<0$ and low density in DORI-$\lambda_2$ analysis, where the average NMAPE is 0.42 and the absolute error magnitude is MAE = 0.0031 e/Å³, while the average NMAPE is 0.35 for intralayer covalent bonding regions ($\lambda_2$ <0, high density). These results indicate that even without explicit training on interlayer interaction data, DeePAW still can accurately capture the characteristics of electron density under weak interactions through the coupling relationship between the curvature and density fluctuations in the van der Waals interaction regions of the DORI-$\lambda_2$ plane. DeePAW has the strong generalization capacity, robust and accurate predictive ability on the symmetry-broken bilayer graphene. Furthermore, armchair and zigzag carbon nanotubes (CNTs) are selected as the typical 1D materials to illustrate the power in the electron density and formation energy prediction of DeePAW, where the armchair and zigzag CNT are denoted by (20, 20) and (20, 0), respectively. In the single-unit periodic structure, the armchair CNT has a diameter of 2.71 nm and length of 0.12 nm, containing 80 atoms, while the zigzag CNT has a diameter of 1.57 nm and length of 0.28 nm, containing 80 atoms. Again, the KSDFT calculations are carried out to evaluate the prediction accuracy. Besides indicating the formation energy, Fig. 4e shows the projected view of electron density predicted by DeePAW and calculated by KSDFT, which gives the values of NMAPE=0.56 and $R^2$ = 0.9999 for the armchair CNT and NMAPE=0.57 and $R^2$ = 0.9999 for the zigzag CNT.

**Performance of DeePAW on Catalysis and Light Absorption (Emission)**

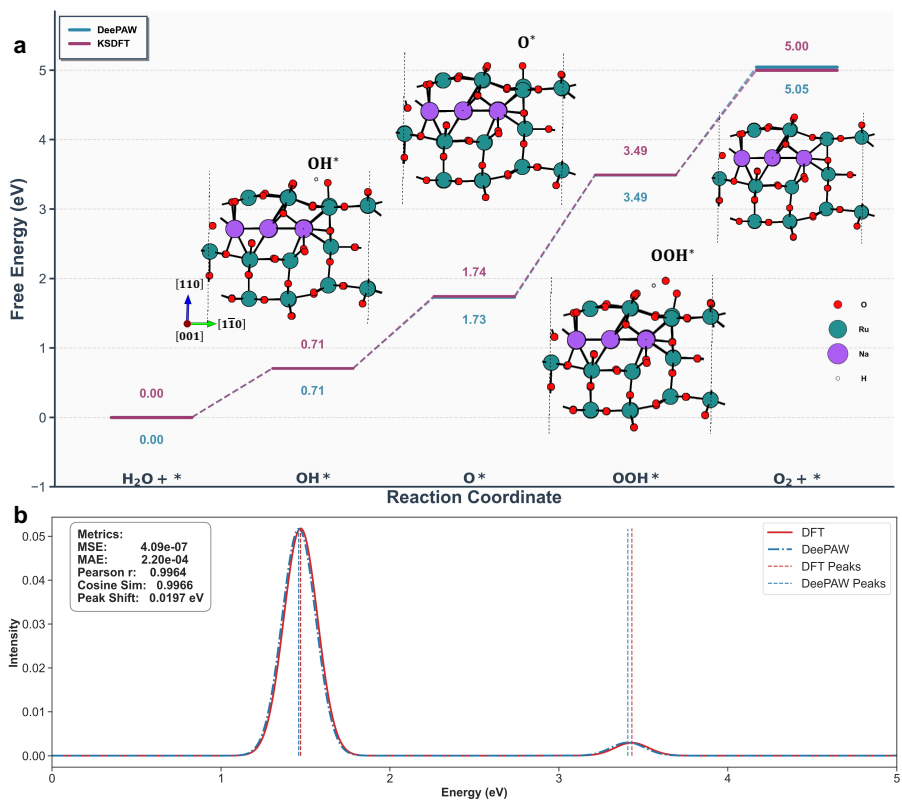

**Figure 5 a**. The oxygen evolution reaction pathway on the (110) surface of tetragonal Na-RuO$_2$ catalyst, DeePAW predicted and KSDFT calculated, indicates that the energy difference for the rate-determining step (RDS) in the OER reaction pathway is only 0.05 eV. **b.** The absorption spectrum of the LiBiO$_3$ system, TD-DeePAW predicted and TD-KSDFT calculated.

Recently, Qin et al.[47] reported the excellent catalysis capability of Na doped rutile RuO$_2$ (space group 136, P4_2/mnm) crystals in oxygen evolution reaction (OER) in acidic water electrolyte. With the predictive capacity of formation energy, the OER pathway is predicted by DeePAW and calculated by KSDFT for comparison (see details in the Supplementary Materials). Fig. **5a** demonstrates the predictive catalysis capability on the (110) plane of the Na-RuO$_2$ system, the predicted energy at the rate-determining step (RDS) in the OER reaction pathway is in excellent agreement with the KSDFT calculated, evidenced by the maximal difference of 0.05eV between both. More importantly, differential electron density analysis of the RDS reveals that the electron redistribution patterns predicted by DeePAW closely match those from KSDFT calculated in both spatial distribution features and magnitude. The regions of electron transfer induced by oxygen intermediate adsorption, as well as the fine structures of charge accumulation and depletion near the transition state, are accurately reproduced by DeePAW. This demonstrates that DeePAW not only achieves high accuracy in energetics but also faithfully captures the essential electronic structural changes during catalytic processes.

The time-dependent (TD) DFT theory[48,49] handles the dynamical perturbation to the ground state electron density field, which causes excited states as the response of the ground state electron density to a time-varying applied electric field. These excitation energies are linked to the called Bohr frequencies with Planck constant[50]. Here, DeePAW is extended to predict optical properties of structures via the open code[51,52] (https://gitlab.com/pavanello-research-group/dftpy), which workflow is briefly described in the Supplementary Information. We select the cubic perovskite LiBiO$_3$ (space group 56, P4mm, lattice parameters a = b = c = 3.82 Å) from the dataset (see Methods for detail) and calculate its light adsorption spectrum. Fig. **5b** shows the absorption spectrum of the LiBiO$_3$ system[41]. The low-energy excitation corresponds to the first peak primarily arises from collective electronic oscillations in the bonding region, closely related to the ground-state electron density distribution and its gradient in this region. These oscillations simultaneously influence both the excitation energy (frequency) and the oscillator strength. For the first peak, the TD-KSDFT calculated yields an excitation energy of 1.4735 eV, while the TD-DeePAW predicts 1.4575 eV, resulting in a small error of -0.016 eV (relative deviation of -1.09%). Regarding oscillator strength, TD-KSDFT gives a value of $5.18 \times 10^{-2}$, and TD-DeePAW predicts $5.17 \times 10^{-2}$, with a relative deviation of only -0.22%. The excellent agreement in oscillator strength not only validates the accuracy of the predicted excitation energies but, more importantly, further confirms that the model perfectly reproduces the integral characteristics of the spatial distribution and gradient of the electron density, particularly capturing the dynamic behavior of electron density and its gradient variations in critical regions such as near atomic nuclei and within bonding zones. The second peak, corresponding to higher-energy excitations, is located in the transition region between the bonding electron zone and the vicinity of atomic nuclei, with its physical origin involving coupled excitations from both regions. TD-KSDFT calculates the excitation energy as 3.4354 eV, while TD-DeePAW predicts 3.4114 eV, yielding a deviation of -0.024 eV (relative deviation of -0.70%). For oscillator strength, TD-KSDFT reports $2.94 \times 10^{-3}$, and TD-DeePAW predicts $3.00 \times 10^{-3}$, with a deviation of +1.94%. In this region, the electron density gradient undergoes a transition from gentle to steep, requiring the model to handle features across multiple spatial scales. The consistency in both frequency and oscillator strength further demonstrates the model's stability across different length scales and highlights its strong capability for multiscale description. From the overall spectral distribution, the Kullback-Leibler (KL) divergence is only 0.0029, which implies that the two distributions are nearly identical. These results reveal that TD-DeePAW can master the underlying spatial structure and predict the collective excitation properties with such accuracy approaching that of TD-KSDFT.

**Discussion**

An excellent universal machine learning model for OFDFT calculations must possess three outstanding characteristics: the as larger as possible coverage of elements, the as wide as possible diversity of applications, and the as accurate as possible predictions. Guided by this tripartite criterion, DeePAW represents a paradigm shift from specialized chemical-specific models toward a universal model for electron-based materials science, yet critical challenges are still faced for DeePAW to become a truly "universal" electronic structure solver.

As mentioned above, DeePAW covers 88 elements in the MP database and its predictions of formation energy and electron density are all with considerable accuracy on the 117,452 crystal structures involving the 88 elements. Obviously, if a database includes all elements in the periodic table, DeePAW trained on that database will cover all elements. The outstanding performance is attributed to the model architecture of double MPNNs and three output heads. DeePAW demonstrates that integrating physical priors into deep learning architectures is key to handling chemical diversity. The atomic MPNN feeds the atomic embeddings, layer by layer, to the electron density MPNN. The first training of electron density functional also renders the atomic MPNN certain electronic features and thus it can accurately predict properties at the structure level, such as the formation energy. In the electron density prediction, DeePAW partitions the electron density functional into smooth and fluctuating components—conceptually mirroring the Projector Augmented-Wave (PAW) formalism—the model effectively decouples the learning of delocalized valence electrons from localized core oscillations, which are separately handled by the two output heads, MLPs for smooth fields and KANs for rapid local fluctuations. In this sense, the two heads behave like a ResNet[53] because the loss of KANs utilizes the whole information of labels. The excellent performance of DeePAW on electron density is evidenced by a state-of-the-art NMAPE of 0.351% across 88 elements. This significantly outperforms contemporary universal models like ChargE3Net (0.523%) and DeepDFT (0.799%), confirming that architectural inductive biases are essential for maintaining high fidelity across the nearly entire periodic table.

Although trained exclusively on the 3D bulk crystals in MP database, DeePAW exhibits remarkable zero-shot generalization power to other unseen 3D perfect crystals in other datasets, 3D point-defected crystals, 2D monolayers, twisted bilayer graphene and 1D carbon nanotubes without any fine-tuning. This transferability suggests the model has learned the intrinsic local chemistry rather than merely memorizing bulk periodicities. Furthermore, DeePAW bridges the gap between scalar density prediction and complex material functionalities. From replicating the polarization switching pathways in ferroelectric $HfO_2$ to predicting oxygen evolution reaction barriers in $Na-RuO_2$ catalysts with chemical accuracy (0.01 eV deviation), the model proves that accurate density topology is a robust descriptor for diverse physical observables. As a natural extension, TD-DeePAW exhibits the impressive prediction capacity in studying the optical properties of structures.

Despite these advances, the "universality" of DeePAW still faces intrinsic challenges imposed by the database, which in turns on the challenge faced in KSDFT calculations. The fundamental challenge come from tightly correlated electron systems. DeePAW exhibits elevated errors (NMAPE ~10%) for f-electron systems such as U and Pu, highlighting the difficulty of capturing strong electron correlation and spin-orbit coupling effects within KSDFT calculations. The MP database accumulates a large number of KSDFT calculation results over decades, these KSDFT data quality might call for careful assessments, with regarding to convergence criteria, pseudopotential versions, etc. Nevertheless, the performance of DeePAW is robust and excellent on main-group and transition elements. The current version DeePAW predicts only formation energy and electron density field and the next version DeePAW will consider to predict atomic forces or stress tensors, aiming to direct geometry optimization and ab initio molecular dynamics (MD) without additional gradients derived from the potential energy surface.

## Methods

### Datasets

The used dataset to train, validate and test DeePAW comes from the Materials Project (MP) database that contains 117,452 crystal structures and covers 90 elements, spanning all seven crystal systems, 14 Bravais crystal lattices and 212 space groups. The two elements of He and Ne are only participated in three crystals and thus removed out, thereby leaving 88 elements. The 117,452 crystal structures are randomly divided into the training and testing subsets at the ratio 9:1 and nine-fold cross-validation is conducted with the training subset. The following additional datasets are used to access the prediction performance of pretrained DeePAW. The cubic 3D structure LaGaO$_3$ ($221, Pm\bar{3}m$) is from the perovskite-5 dataset[41]. The 2x2x1 superlattice silicon including one Ba atom at the interstitial site (217 atoms in total) is copied from entry-424 in MPcontribs dataset [42]. The 2D CsF structure 2D crystal structures from the Computational 2D Materials Database (C2DB)[54]. The twisted graphene dataset is provided by the DeepH open-source project[20]. The carbon nanotube structures are selected from the dataset[55], including both armchair chirality and zigzag chirality.

### KSDFT calculations

To verify the predictions from DeePAW, KSDFT calculations are carried out by using the Vienna Ab initio Simulation Package (VASP version 6.4.3). The exchange-correlation interactions are modeled using the generalized gradient approximation (GGA) with the Perdew-Burke-Ernzerhof (PBE) functional. The convergency criterion for the electronic self-consistent field (SCF) iterations is set as that the energy difference between two successive calculation steps is smaller or equal to $1\times 10^{-7}$ eV for the calculated superlattice.

### Grid Construction and Labels of Electron Density

The VASP coarse-fine grid scheme[37] is used here to construct electron density grids and a more detailed description is provided in Supplementary Information. Following the Nyquist-Shannon sampling theorem, the number of the coarse grid points along lattice direction $i$ must satisfy $NG_i \geq \frac{G_{cut}^{eff} \times |\vec{a}_i|}{2}$, where $|\vec{a}_i|$ is the magnitude of the $i$-th lattice vector and $G_{cut}^{eff}$ denotes the effective cutoff wavevector, which is related to the cutoff wavevector $G_{cut}$ via $G_{cut}^{eff} = C_{PREC} \times G_{cut}$ with $C_{PREC}$ being a preset parameter[56] (see Supplementary Information for detail) and $C_{PREC} = 1.3$ used here. The cutoff wavevector $G_{cut}$ is calculated from the plane wave kinetic energy cutoff $G_{cut} = \sqrt{\frac{2m}{\hbar^2} \cdot ENCUT}$, where m is the electron mass and $\hbar$ is the Plank constant so that the coefficient $\frac{2m}{\hbar^2} = 0.262465\ \text{Å}^{-2}eV^{-1}$. The number of fine grid points along lattice direction $i$ is twice of the number of coarse grid points so that $NGF_i \geq G_{cut}^{eff} \times |\vec{a}_i|$. This work takes ENCUT=520 eV[5] and has $G_{cut}^{eff} \approx 15.18\ \text{Å}^{-1}$, which gives the number of linear grid points $NGF_i \approx 15\ |\vec{a}_i|/\text{Å}$. Usually, the grid points are uniformly distributed within the lattice cell. Obviously, the larger the lattice cell is, the more number the grid points will be.

The labels of electron density are calculated from orbital wave functions $\Psi_i(r)(i = 1,\cdots)$ obtained by KSDFT at point $r$, viz., $\rho(\boldsymbol{r}) = \sum_i |\Psi_i(\boldsymbol{r})|^2$.

**Model Design**

The DeePAW model employs double E(3)-invariant MPNNs, with the periodic boundary conditions, that achieve both electron density and formation energy predictions through end-to-end automated representation learning. The present work uses message functions to updates node embeddings layer by layer and finally delivers node embeddings to the readout phase of output head. With the atomic configuration, the atomic presentation $Z_z \in \mathbb{R}^{1 \times 118}$ $(z = 1,\cdots,118)$ denotes the atomic number or pseudo-nucleus number of atom $z$ and expressed by a one-hot vector to cover all the 118 elements in the periodic table. The initial embedding of atom $i$ in a structure is given by

$$\mathbf{h}_i^{(0)} = \text{embedding}\ (\Theta_z, Z_{iz})\ (z = 1,\cdots,118)(i = 1,\cdots,N_a), \tag{1}$$

where $\Theta_Z$ denotes all the involved trainable parameters and $\mathbf{h}_i^{(0)} \in \mathbb{R}^{1 \times d}$ with d being the dimension of node embeddings in the MPNN. The edge embeddings $e_s(|r_{ij}|) \in \mathbb{R}^{1 \times S}$ is constructed by Gaussian basis function (GBF)

$$e_s(|r_{ij}|) = \frac{1}{\sigma\sqrt{2\pi}} \exp\left(-\frac{(|r_{ij}|-r_s)^2}{2\sigma^2}\right), (s = 1,2,\cdots,S), \quad (2)$$

where $|r_{ij}|$ is the distance between node i and node j, $\sigma^2 = 0.0044$ is the variance, and scalar $r_s$ $(s = 1, \cdots, S)$ is placed uniformly with an identical step between 0 and 6 Å. Eq. (2) indicates that initial edge embedding takes the real distance as the mean, sets a hyperparameter as variance, and then sample S values from the Gaussian distribution $N(|r_{ij}|, \sigma^2)$. After initiation, the message functions $\mathbf{m}_{ij}^{(n)}$ are recursively calculated by

$$\mathbf{m}_{ij}^{(n)} = \sum_{l_1,l_2} \sum_{m_1,m_2} C_{l_1 m_1, l_2 m_2}^{LM} \cdot Y_{l_1 m_1}(\theta_{i,j}, \varphi_{i,j}) \cdot h_j^{(n-1)} \cdot \sigma\left(\sum_s e_s(|r_{ij}|) \cdot W_{l_1 l_2 s}\right), (n = 1, \cdots, N_L), \quad (3a)$$

where $C_{l_1 m_1, l_2 m_2}^{LM}$ denotes the Clebsch-Gordan coefficients, $Y_{l_1 m_1}(\theta_{i,j}, \varphi_{i,j})$ is the spherical harmonics, $h_j^{(n-1)}$ is the node j embeddings of layer $(n-1)$, and $\sigma(\cdot)$ is the Sigmoid Linear Unit (SiLU) activation function, and $W_{l_1 l_2 s}$ is a learnable parameter that weights the combination of angular momentum components. Here, $\theta_{i,j}$ and $\varphi_{i,j}$ denote the polar and azimuthal angles, respectively, that define the direction from atom $i$ to its neighbor $j$. When processing scalar features $(l = 0)$, the spherical harmonics degenerate to a constant $Y_{00} = \frac{1}{\sqrt{4\pi}}$, and the tensor product operation reduces to scalar multiplication,

$$\mathbf{m}_{ij}^{(n)} = h_j^{(n-1)} \cdot \sigma\left(\sum_s e_s(r_{ij}) \cdot W_{00s}\right) (n = 1, \cdots, N_L). \quad (3b)$$

Each atom $i$ aggregates messages from all its neighbors $j \in \mathcal{N}(i)$

$$\mathbf{M}_i^{(n)} = \frac{1}{\sqrt{|\mathcal{N}(i)|}} \sum_{j \in \mathcal{N}(i)} \mathbf{m}_{ij}^{(n)} \tag{4}$$

The aggregated message $\mathbf{M}_i^{(n)}$ is used to up date the node embeddings

$$\mathbf{h}_i^{(n)} = \cos\phi_i \cdot \mathbf{h}_i^{(n-1)} + \sin\phi_i \cdot g\left(\mathbf{M}_i^{(n)}\right), \tag{5}$$

where $\phi_i$ represents the dynamic rotation angle of atom $i$ and $g(\cdot)$ originates from the gating mechanism in *e3nn* (see Supplementary Information for detail). The update of node embeddings dynamically adjusts the mixing ratio between the input $\mathbf{h}_i^{(n-1)}$ and the residual component $g(\mathbf{M}_i^{(n)})$.

In electron density MPNN, the initial electron density of node k, $\tilde{\mathbf{h}}_k^{(0)}(k = 1, \cdots, N_{ED})$, might be homogeneously or randomly set-up under the constraints $\rho(\mathbf{r}) \geq 0$ and $N_e = \sum_i^{N_\rho} \rho(\mathbf{r}_i)$. The edge embeddings $\tilde{e}_s(|\tilde{r}_{ik}|) \in \mathbb{R}^{1 \times S}$ is also constructed by Gaussian basis function (GBF) except that $|\tilde{r}_{ik}|$ denotes the distance from grid point k to its neighboring nucleus i, viz.,

$$\tilde{e}_s(|r_{ik}|) = \frac{1}{\sigma\sqrt{2\pi}} \exp\left(-\frac{(|r_{ik}|-r_s)^2}{2\sigma^2}\right), (s = 1, 2, \cdots, S). \tag{6}$$

The node embeddings in the atomic MPNN will be fed into the ED MPNN and the message functions $\widetilde{\mathbf{m}}_{ij}^{(n)}$

$$\widetilde{\mathbf{m}}_{ik}^{(n)} = \sum_{l_1,l_2} \sum_{m_1,m_2} C_{l_1 m_1, l_2 m_2}^{LM} \cdot h_i^{(n)} \cdot \tilde{Y}_{l_2 m_2}(\tilde{\theta}_{i,k}, \tilde{\varphi}_{i,k}) \cdot \sigma(\tilde{e}_s(|\tilde{r}_{ik}|) \cdot \widetilde{W}_{l_1 l_2 s}) \ (n = 1, \cdots, N_L). \tag{7a}$$

Similarly, each grid point $k$ aggregates messages from all information related to its neighboring nuclei $i \in \mathcal{N}(k)$ by

$$\widetilde{\mathbf{M}}_k^{(n)} = \frac{1}{\sqrt{|\mathcal{N}(k)|}} \sum_{i \in \mathcal{N}(k)} \widetilde{\mathbf{m}}_{ik}^{(n)}. \tag{7b}$$

Then, the embeddings of grid points are updated by

$$\tilde{\mathbf{h}}_k^{(n)} = \mathbf{w}_{\text{share}}^{(n)} \cdot \cos\phi_k \cdot \tilde{\mathbf{h}}_k^{(n-1)} + \mathbf{w}_{\text{inter}}^{(n)} \cdot \sin\phi_k \cdot \widetilde{\mathbf{M}}_k^{(n)} \ (n = 1, \cdots, N_L), \tag{7c}$$

where $\mathbf{w}_{\text{share}}$ and $\mathbf{w}_{\text{inter}}$ denote the shared weight vector and interaction weight vector, respectively. These vectors are used to control the relative contributions of the node's own features from the previous layer and the environmental information during the message passing process (see Supplementary Information for detail).

The node embeddings from all layers in the atomic MPNN are fed into the GAT head, which delivers the formation energy of a crystal structure (see Supplementary Information for detail).

$$E = \text{GAT}\left(\Theta_{GAT}, h_i^n(v_i)\right) \ (n = 1, \cdots, N_L) \ and \ (i = 1, \cdots, N_a). \tag{8}$$

The node embeddings from all layers in the ED MPNN are fed into the MLP head, which delivers the smoothly varied electron density profile (see Supplementary Information for detail)

$$\tilde{\rho}^{MLP}(r_i) = \text{MLP}\left(\Theta_{MLP}, \tilde{\mathbf{h}}_i^n(r_i)\right) \ (n = 1, \cdots, N_L). \tag{9}$$

The node embeddings of the last layer in the ED MPNN are fed into the KAN head to calculate residual electron density at all grid points, viz.,

$$\Delta\tilde{\rho}(r_i) = \text{KAN}\left(\Theta_{KAN}, \tilde{\mathbf{h}}_i^{N_L}(r_i)\right). \tag{10}$$

In Eqs. (8-10), $\Theta_{GAT}$, $\Theta_{MLP}$ and $\Theta_{KAN}$ represent all corresponding trainable parameters.

**Model Training**

The MP training dataset of 154,719 structures is used for training and nine-fold cross-validation. The training process is divided into two phases and loss functions of L1 loss are used for the MLP, KAN and GAT, respectively. The first training phase optimizes all trainable parameters in the double MPNNs, the MLP head and the KAN head, and the pretrained DeePAW can predict electron density functional. After that, the optimal trainable parameters in the double MPNNs are fixed and the second training phase optimizes only these trainable parameters in the GAT head. More detailed descriptions of training and validation are provided in the Supplementary Information.

**Density Overlap Region Indicator and $\lambda_2$ Analysis**

DORI (Density Overlap Region Indicator) is defined by[57]

$$\text{DORI} = \frac{\phi}{1+\phi} \; with \; \phi = \frac{\left|\nabla\left(\frac{|\nabla\rho|}{\rho}\right)^2\right|^2}{\left(\frac{|\nabla\rho|}{\rho}\right)^6}. \tag{11}$$

The $\lambda_2$ value is the intermediate eigenvalue of the electron density Hessian matrix, with physical significance closely related to the second derivative (curvature) of the electron density functional[58]. Combining both $\lambda_2$ and DORI, electron density distribution landscape can be examined properly (see Supplementary Information for detail).